\documentclass[]{emulateapj}

\usepackage{apjfonts}

\def\LCDM{$\Lambda\mbox{CDM}$}
\def\Msun{\hbox{$\rm\thinspace M_{\odot}$}}
\def\Mpc{{\rm\thinspace Mpc}}
\def\yr{{\rm\thinspace yr}}  
\def\Myr{{\rm\thinspace Myr}}

\def\Msunyr{{\Msun\yr^{-1}}}

\shorttitle{The Millennium Simulation Compared to $z\approx2$ Galaxies}
\shortauthors{Genel et al.}

\begin{document}

\title{Mergers and Mass Accretion Rates in Galaxy Assembly: The Millennium Simulation Compared to Observations of $\lowercase{z}\approx2$ Galaxies}

\author{Shy Genel\altaffilmark{1}, Reinhard Genzel\altaffilmark{1,2}, Nicolas Bouch{\'e}\altaffilmark{1}, Amiel Sternberg\altaffilmark{3}, Thorsten Naab\altaffilmark{4}, Natascha M. F\"orster Schreiber\altaffilmark{1}, Kristen L. Shapiro\altaffilmark{5}, Linda J. Tacconi\altaffilmark{1}, Dieter Lutz\altaffilmark{1}, Giovanni Cresci\altaffilmark{1}, Peter Buschkamp\altaffilmark{1}, Richard I. Davies\altaffilmark{1}, Erin K. S. Hicks\altaffilmark{1}}

\altaffiltext{1}{Max Planck Institut f\" ur extraterrestrische Physik, Giessenbachstrasse, D-85748 Garching, Germany; shy@mpe.mpg.de; genzel@mpe.mpg.de; nbouche@mpe.mpg.de; forster@mpe.mpg.de; linda@mpe.mpg.de; lutz@mpe.mpg.de; gcresci@mpe.mpg.de; buschkamp@mpe.mpg.de; davies@mpe.mpg.de; ehicks@mpe.mpg.de.}
\altaffiltext{2}{Department of Physics, Le Conte Hall, University of California, Berkeley, CA 94720.}
\altaffiltext{3}{School of Physics and Astronomy, Tel Aviv University, Tel Aviv 69978, Israel; amiel@wise.tau.ac.il.}
\altaffiltext{4}{Universit\"ats-Sternwarte M\"unchen, Scheinerstr.\ 1, D-81679 M\"unchen, Germany; naab@usm.uni-muenchen.de.}
\altaffiltext{5}{Department of Astronomy, Campbell Hall, University of California, Berkeley, CA 94720; shapiro@astron.berkeley.edu.}

\slugcomment{Accepted for publication in ApJ}

\begin{abstract}
Recent observations of UV-/optically selected, massive star forming galaxies at $z\approx2$ indicate that the baryonic mass assembly and star formation history is dominated by continuous rapid accretion of gas and internal secular evolution, rather than by major mergers. We use the Millennium Simulation to build new halo merger trees, and extract halo merger fractions and mass accretion rates. We find that even for halos not undergoing major mergers the mass accretion rates are plausibly sufficient to account for the high star formation rates observed in $z\approx2$ disks. On the other hand, the fraction of major mergers in the Millennium Simulation is sufficient to account for the number counts of submillimeter galaxies (SMGs), in support of observational evidence that these are major mergers. When following the fate of these two populations in the Millennium Simulation to $z=0$, we find that subsequent mergers are not frequent enough to convert all $z\approx2$ turbulent disks into elliptical galaxies at $z=0$. Similarly, mergers cannot transform the compact SMGs/red sequence galaxies at $z\approx2$ into observed massive cluster ellipticals at $z=0$. We argue therefore, that secular and internal evolution must play an important role in the evolution of a significant fraction of $z\approx2$ UV-/optically and submillimeter selected galaxy populations.
\end{abstract}


\section{Introduction}
\label{s:intro}
In the cold dark matter model of hierarchical structure formation \citep{BlumenthalG_84a,DavisM_85a,SpringelV_06a} mergers are believed to play an important role in galaxy formation and evolution \citep{SteinmetzM_02a}. Mergers induce starbursts \citep{HernquistL_95a} and transform galactic morphology \citep{NaabT_03a}. Major mergers may drive the buildup of the red sequence \citep{ToomreA_77a,HopkinsP_07b}. Dark matter models and many observations show that mergers are more frequent at high redshift \citep{FakhouriO_07a,ConseliceC_03b}.

However, there is growing evidence that a smoother growth mode may be important for the baryonic mass assembly and star formation history at high redshift. For example, the tight correlation between star formation rate (SFR) and stellar mass in UV-/optically selected star forming galaxies is indicative of buildup by continuous gas inflow \citep{DaddiE_07a,NoeskeK_07a}. As part of the SINS survey (\citealp{FoersterSchreiberN_06a}; N.~M.~F\"orster Schreiber et al.~2008 in preparation), integral field spectroscopy of more than $50$ UV-/optically selected $z\approx2$ star forming galaxies show a preponderance of thick gas-rich rotating disks and only a minority of major mergers \citep{FoersterSchreiberN_06a,GenzelR_06a,GenzelR_08a,ShapiroK_08a}. In contrast, SMGs are probably short-lived maximum-starburst galaxies undergoing dissipative major mergers \citep{TacconiL_06a,BoucheN_07a,TacconiL_08a}. Table \ref{t:galaxies} summarises key properties of these $z\approx2$ galaxy samples.

How do these observations fit into the concordance cosmological model? Modern simulations of dark matter structure formation are robust and fixed by the cosmological parameters. However, complicated baryonic physics makes it difficult to model the evolution of galaxies and reproduce, for example, the high SFRs of these $z\approx2$ galaxies \citep{DaddiE_07a}.

Galaxies at $z\approx2$ differ significantly from local galaxies. The central mass densities of SMGs and of massive quiescent galaxies at the same redshift (\citealp{vanDokkumP_07a} and references therein) are an order of magnitude greater than those of local spheroids and disks \citep{TacconiL_08a}. Also, the $z\approx2$ rotating disks are thick and turbulent, unlike local disk galaxies. These differences raise the question: what are the local Universe descendants of these $z\approx2$ galaxies?

In this paper we use the cosmological dark matter Millennium Simulation (\citealp{SpringelV_05a}; \S\ref{s:find_merger}) to investigate the possible role of major mergers in galaxy formation at $z\approx2$ (\S\ref{s:z_2}), and to consider the evolution of the $z\approx2$ galaxies to $z=0$ (\S\ref{s:evolution}).

\begin{table}[tbp]
\caption{Properties of galaxy samples at $z\approx2$}
\label{t:galaxies}
\centering          
\begin{tabular}{c| c c c c} 
\hline\hline       
Galaxy & SFR & Halo & Comoving number & Major \\ 
sample & $[\Msunyr]$ & mass & density & merger \\ 
&& $[\Msun]$ &[$h_{0.7}^3\Mpc^{-3}$]& fraction \\
\hline                    
SINS & $\approx30-300$\tablenotemark{(a)} & $10^{11.84}v_{200}^3\times$ & $1-2.2\times10^{-4}$\tablenotemark{(c,d)} & $\approx0.3$\tablenotemark{(e)} \\
&&$(\frac{1+z}{3.2})^{-1.5}h_{0.7}^{-1}$\tablenotemark{(b)}&&\\
SMGs\tablenotemark{(f)} & $\approx750\pm300$\tablenotemark{(g)} & - & $1-2\times10^{-5}$\tablenotemark{(c)} & $\approx1$\tablenotemark{(c)} \\
\hline                  
\end{tabular}
\tablecomments{(a) N.~M.~F\"orster Schreiber et al.~2008 in preparation ; (b) \citealp{FoersterSchreiberN_06a} ; (c) \citealp{TacconiL_08a} and references therein ; (d) BX/BM \& sBzK galaxies with $K\leq20$ ; (e) \citealp{ShapiroK_08a} ; (f) $S(850{\rm \mu m})\geq5{\rm mJy}$  ; (g) Genzel, priv.~comm.}
\vspace{0.5cm}
\end{table}

\section{Analysis of the Millennium Simulation}
\label{s:find_merger}

\subsection{The Millennium Simulation and its merger trees}
\label{s:millennium}
The Millennium Simulation is a cosmological N-body simulation. It follows $2160^3$ dark matter particles of mass $8.6\times10^8h^{-1}\Msun$ in a box of $500h^{-1}\Mpc$ on a side from $z=127$ to $z=0$. There are 64 output times (``snapshots''), at $\approx300\Myr$ intervals at $z\lesssim3$. The cosmology is \LCDM, with $\Omega_m=0.25$, $\Omega_{\Lambda}=0.75$, $\Omega_b=0.045$, $h=0.73$, $n=1$ and $\sigma_8=0.9$.

In the Millennium Simulation, structures are identified in two steps. First, the Friends-of-Friends (FOF) algorithm \citep{DavisM_85a} creates a catalogue of FOF groups at each snapshot. The FOF groups represent dark matter halos. Second, bound substructures are identified inside the FOF groups (SUBFIND; \citealp{SpringelV_01}), so that each halo contains at least one subhalo. The Millennium merger trees are constructed by finding a single descendant for each {\it subhalo} in the following snapshot, while the FOF groups themselves play no role in constructing the merger trees.

In traditional merger trees, mergers are instantaneous, i.e. there is no information on their durations. Therefore, the Millennium public merger trees\footnote{The Structure catalogues and derived merger trees have been made public by the Virgo Consortium: http://www.mpa-garching.mpg.de/millennium.} can be used to determine the merger rate, which is merely a count of the number of mergers per unit time (\citealp{FakhouriO_07a}, S.~Genel et al.~2008 in preparation). However, they cannot be used to determine important quantities such as the major merger fraction, defined as the fraction of halos undergoing major mergers at a given time, or the mass growth rate of each halo.

\subsection{Constructing new merger trees}
\label{s:new_trees}
To derive merger fractions and mass growth rates, we must consider the finite physical durations of mergers. Therefore, start and end points must be defined. Also, to derive these quantities for entire dark matter halos rather than for subhalos, new trees have to be constructed based on FOF groups. In our procedure, the main subhalo in each FOF group is identified and is then followed to its subhalo descendant, using the original subhalo-based trees.  The FOF group to which the subhalo descendant belongs is defined as the FOF group descendant of the original FOF group. Thus, in our new trees each node is an entire FOF group, rather than a subhalo. If two subhalos merge while within a single FOF group we do not count this as a merger event.

We identify a merger whenever two or more FOF groups at snapshot $s$ have a common descendant at snapshot $s+1$. However, at this time the halos are not necessarily already physically merging, since they may still be well separated. To account for that, we track the distances between the subhalo descendants of the main subhalos of the original FOF groups. These subhalos represent, approximately, the centers of the entire halos. We then define the start point of the merger as the last snapshot at which this distance is still larger than the sum of the virial radii of the original halos (FOF groups). In some mergers the subhalos disappear before the distances become smaller than the sum of the original virial radii. When this happens the start point is defined as just one snapshot prior to the point where the subhalos merge and/or disappear.

After a merger begins, one of the halos becomes a substructure within the other. This substructure typically dissolves too quickly to be followed until the merger is physically complete. To overcome this problem we first estimate the duration of mergers $T_{\rm merger}$, and define their end point as $T_{\rm merger}$ after the start point. To estimate the durations, we considered the fitting functions of \citet{Boylan-KolchinM_07a} and \citet{JiangC_07a}, which are based on simulations of mergers. Our results are qualitatively robust with respect to this choice. We present quantitative results based on the orbit-averaged \citet{Boylan-KolchinM_07a} fitting function for the dynamical friction merger time: $T_{\rm merger}=0.05\frac{r^{1.3}}{ln(1+r)}\frac{1}{H(z)}$, where $r$ is the mass ratio and $H(z)$ is the Hubble constant at redshift $z$.

The accretion rate we associate with each merger equals the amount of accreted mass divided by the merger duration $T_{\rm merger}$.

To summarise, we construct new FOF group-based merger trees, where each FOF group also holds information about internal on-going mergers. The accretion rates associated with those mergers are summed up to obtain the total accretion rate onto the FOF group in question. The merger mass ratio is determined by the masses of the FOF groups at snapshot $s$, just prior to the appearance of a common FOF group descendant at snapshot $s+1$. We define major mergers as those with mass ratios between $3:1$ and $1:1$ (with $1:1$ being the most ``intense'' type of merger). If the most intense merger associated with a halo lies between $3:1$ and $1:1$, the halo is labelled as undergoing a major merger.

\section{Galaxies at $\lowercase{z}\approx2$}
\label{s:z_2}
\begin{figure}[t]
\centering
\includegraphics[]{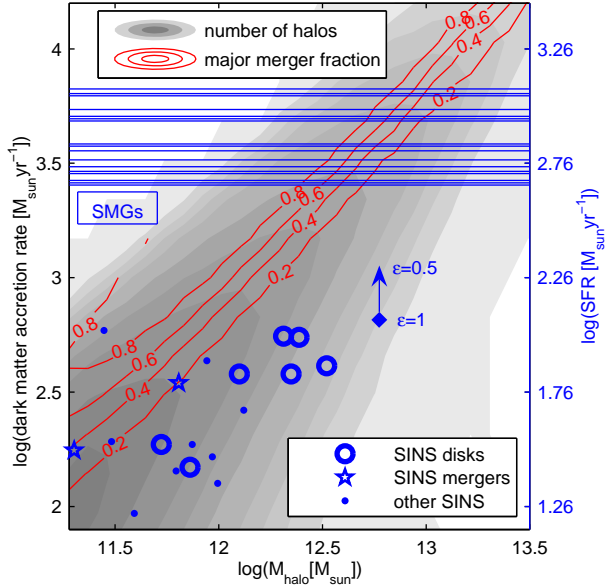}
\caption{The distribution ({\it shaded contours}) of $z\approx2.2$ halos in the dark matter accretion rate (left y-axis) versus halo mass plane. Also major merger fractions are displayed ({\it red contours}). Associated SFRs are indicated (right y-axis) assuming an effective star formation efficiency $\epsilon=1$ in eq.~\ref{e:SFR_DAR}. SINS galaxies at $2<z<2.5$ are indicated as disks ({\it circles}) or mergers ({\it stars}) based on the \citet{ShapiroK_08a} classification, and others ({\it points}) not investigated by \citet{ShapiroK_08a}. Their SFRs are based on their $H\alpha$ fluxes corrected for extinction using $A_{H\alpha}=0.8$. Their halo masses were determined by assuming that the observed disk maximum rotation velocity is equal to the circular velocity of the halo \citep{FoersterSchreiberN_06a}, and only galaxies with $M_{\rm halo}>10^{11.25}\Msun$ are included. The SFRs for the SMGs are shown as the horizontal lines in the upper part of the figure (because the SMGs are compact, their halo masses cannot be reliably inferred from the observed gas motions). The {\it left} y-axis indicates what dark matter accretion rates are needed to account for the observed SFRs when assuming $\epsilon=1$. The arrow indicates the shift in the galaxy positions if $\epsilon=0.5$.}
\vspace{0.5cm}
\label{f:f1}
\end{figure}

Fig.~\ref{f:f1} shows halo number densities ({\it shaded contours}) and major merger fractions ({\it red contours}) as functions of halo mass and dark matter accretion rate for $z\approx2.2$. It shows that the major merger fraction is an increasing function of {\it specific} dark matter accretion rate ($\frac{\dot{M}_{\rm DM}}{M_{\rm halo}}$), as both quantities increase towards the upper-left direction of the plane. This trend holds at all redshifts.

The mean accretion rate scales with halo mass and redshift as

\begin{eqnarray}
\langle\dot{M}_{\rm DM}\rangle\approx35\Msunyr(1+z)^{2.2}M_{12}^{1.07}
\label{e:DAR_M_z}
\end{eqnarray}
(where $M_{12}\equiv\frac{M}{10^{12}\tiny{\Msun}}$). The $1\sigma$ scatter equals $\approx\langle\dot{M}_{\rm DM}\rangle\times(\frac{2.5}{1+z})^{0.2}$, which reflects more the upwards scatter, although negative accretion rates do exist for some halos (because of tidal stripping or fluctuations related to the FOF algorithm). Our numerical results are in good agreement with the analytic approximation for the accretion rate presented by \citet{NeisteinE_06a}, which is based on the extended Press-Schechter (EPS) model \citep{PressW_74a,BondJ_91a,BowerR_91a}. The \citet{NeisteinE_06a} approximation has a somewhat stronger mass and redshift dependence compared to our results. For example, at $z=0$ and $M=10^{12}\Msun$, their accretion rate is $\approx10\%$ higher, and at $z\approx3$ and $M=10^{14}\Msun$ it is a factor of $\approx2$ higher. For halos of particular interest for this paper, i.e.~of $M\approx10^{12}\Msun$ at $z\approx2$, the \citet{NeisteinE_06a} approximation exceeds our eq.~\ref{e:DAR_M_z} by $\approx30\%$.

To compare our results to observed galaxies, we assume that the galaxies are the central galaxies of their halos. We convert dark matter accretion rate ($\dot{M}_{DM}$) into SFR ($\dot{M}_*$) using the baryonic fraction $\eta_B=0.18$ and an effective star formation efficiency $\epsilon$, which is a free parameter used to interpret the results:

\begin{eqnarray}
\dot{M}_*=\eta_B\times\epsilon\times\dot{M}_{DM}.
\label{e:SFR_DAR}
\end{eqnarray}
In the 'cold flow' regime ($M_{\rm halo}\lesssim10^{12}\Msun$; \citealp{BirnboimY_03a,KeresD_05a,OcvirkP_08a}) eq.~\ref{e:SFR_DAR} is a plausible measure of the baryonic accretion rate. At larger masses the accretion rate is lower and is controlled by the cooling time in the hot virialised baryonic halo gas. At much smaller masses it is strongly reduced by outflows generated by supernovae feedback. In the cold flow regime, the cold gas (which may be clumpy) is fed at approximately virial velocity via filaments directly into the halo center, where it accumulates onto the galaxy. Moreover, considering the case where gas is stripped off incoming galaxies, our estimated merger duration is related to the dynamical time of the halo, on which the gas will fall to the central galaxy when the cooling time is short. For major mergers $\epsilon$ may even exceed $1$, because the star formation burst they trigger can be shorter than the dark matter halo merger time scale (e.g.~\citealp{SpringelV_05b,TacconiL_08a}).

\subsection{SINS galaxies}
\label{s:z_2_sins}
On Fig.~\ref{f:f1} we overplot the SINS galaxies. Their SFRs are based on their $H\alpha$ fluxes corrected for extinction using $A_{H\alpha}=0.8$, and their halo masses were estimated by assuming that the observed disk maximum rotation velocity is equal to the circular velocity of the halo \citep{FoersterSchreiberN_06a}. Fig.~\ref{f:f1} shows that if $\epsilon\gtrsim0.5$ is assumed, the host halos of SINS galaxies with $M_{\rm halo}>10^{11.25}\Msun$ lie in the region where most halos of their mass are expected to be concentrated. Furthermore, for $\epsilon\gtrsim0.5$ the expected mass accretion rates are sufficient to account for the observed SFRs. Also, the predicted major merger fraction is small ($\lesssim0.5$), consistent with observations. Although the statistics are still small, we notice that the confirmed SINS mergers (\citealp{ShapiroK_08a}; {\it stars}) have higher specific dark matter accretion rates than the confirmed disks ({\it open circles}), and therefore come from a region where the halo major merger fraction is higher. 

The computed number density of halos with $M\approx10^{11.5}-10^{12}\Msun$ is a few times higher than the observed number density of the galaxies the SINS sample is drawn from (Table~\ref{t:galaxies}). Possibly, the observed galaxies have typically high $M_{\rm gal}/M_{\rm halo}$, with other halos of comparable mass hosting fainter undetected galaxies. Also, some of the halos in this mass range may have already developed virial shocks that quench star formation.

\begin{figure}[t]
\centering
\includegraphics[]{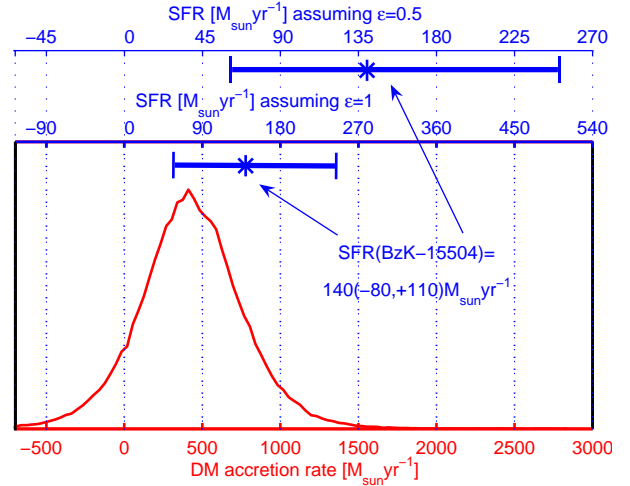}
\caption{The distribution of dark matter accretion rates for halos with masses $(1.2\pm0.3)\times10^{12}\Msun$ at $z\approx2.4$ (corresponding to the halo mass and redshift of BzK-15504), which are not undergoing a major merger. The SFRs on the upper axes are given assuming effective star formation efficiencies $\epsilon=1$ and $\epsilon=0.5$. The measured SFR of $140(-80,+110)\Msunyr$ in BzK-15504 is indicated by the asterisks and error bars.}
\vspace{0.5cm}
\label{f:f2}
\end{figure}

In should be noted that our estimated major merger duration $T_{\rm merger}$ equals $\approx350-1000\Myr$ at $z\approx2.2$, depending on the mass ratio. This is very similar to the "observable" galaxy merger timescale often found in the literature (e.g.~\citealp{ConseliceC_06a,LotzJ_08b}). Therefore, the approximation taken here, i.e.~considering directly the halo merger fraction and inferring the galaxy merger fraction from it, is probably a reasonable one with respect to the deduced merger fraction of the SINS galaxies.

\citet{GenzelR_06a} studied BzK-15504 with high resolution using adaptive optics, and concluded that it was a large proto-disk with no sign of a recent/ongoing major merger, and a SFR of $140(-80,+110)\Msunyr$. \citet{FoersterSchreiberN_06a}, \citet{LawD_07a}, \citet{GenzelR_08a}, G.~Cresci et al.~(2008, in preparation), \citet{BournaudF_08a} and \citet{vanStarkenburg_08a} have found similar systems. Fig.~\ref{f:f2} shows that for halos not undergoing major mergers, and with masses equal to the halo mass of BzK-15504, the typical dark matter accretion rate is $\approx450\Msunyr$, i.e.~the typical SFR assuming $\epsilon=1$ is $\approx80\Msunyr$. About $15\%$ of such halos have SFRs exceeding the $140\Msunyr$ observed in BzK-15504 (again assuming $\epsilon=1$). Thus, the implied dark matter accretion rate in BzK-15504 may be quite typical. Considering the uncertainty in the measured SFR, the implied dark matter accretion rate is consistent with theoretical expectations for $\epsilon$ as low as $\approx0.5$.

We conclude that high star formation rates and large abundances of non-major merger, massive disks at $z\approx2$ are consistent with expectations from \LCDM{ }simulations if accretion is in the 'cold flow' regime and the star formation efficiency is high.

\subsection{SMGs}
\label{s:z_2_smgs}
For $\epsilon\approx1$, the observed SFRs of the SMGs imply dark matter accretion rates of $\approx2500-6000\Msunyr$ (Fig.~\ref{f:f1}). When examining halos with accretion rates in this range that are undergoing major mergers, we find that their masses lie mostly in the range $(2-6)\times10^{12}\Msun$ and obey a log-normal distribution with a mean $\approx3\times10^{12}\Msun$ and $\sigma\approx0.25{\rm dex}$. We also find that their number density is $\approx5\times10^{-5}\Mpc^{-3}$. This is only slightly larger than the observed SMG density (Table~\ref{t:galaxies}), and supports the conclusions of \citet{TacconiL_06a,TacconiL_08a} that the SMGs represent major mergers. If the SMG phase is shorter than the halo merger duration, such that $\epsilon>1$, the implied number density is not much altered, but lower halo masses are found. E.g., if the SMG phase lasts only $100\Myr$, the mean halo mass is $\approx10^{12}\Msun$, in which case SMGs could be members of the UV-/optically selected galaxy populations that have recently experienced a dissipative major merger. The observed rotation velocities of SMGs are larger than the expected circular velocities of halos with these inferred masses. This is consistent with the SMGs being concentrated major mergers where the rotation velocities peak close to the center.

\section{Fate at $\lowercase{z}=0$}
\label{s:evolution}

\subsection{Fate of SINS galaxies}
\label{s:evolution_sins}
Fig.~\ref{f:f3} summarizes the major merger history of halos from $z\approx2.2$ to $z=0$. For halos with initial masses typical of the SINS galaxies' halos, $\approx40\%$ will be accreted via minor mergers by more massive halos (representing groups or clusters) with final masses $10^{13}\Msun\lesssim M\lesssim10^{15}\Msun$ at $z=0$. Around one half of those halos will merge fully with the central subhalo of the group/cluster, and the other half will remain satellite subhalos. The other $\approx60\%$ remain ``main branch'' halos to $z=0$. Of these, $\approx2/3$ undergo at least one major merger during their evolution to $z=0$. Their final halo masses are $10^{12.3}\Msun\lesssim M\lesssim10^{13.3}\Msun$, so their associated galaxies may become massive ellipticals (cf.~\citet{ConroyC_08a}). The other $\approx1/3$ do not undergo any future major mergers, and grow to a mass $10^{11.8}\Msun\lesssim M\lesssim10^{12.5}\Msun$ at $z=0$. Thus, these may evolve via secular evolution into bulges and later possibly grow a new disk.

\subsection{Fate of SMGs}
\label{s:evolution_smgs}

\begin{figure}[t]
\centering
\includegraphics[]{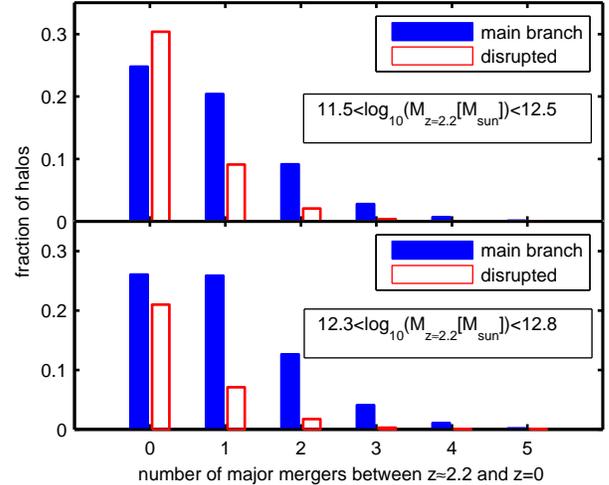}
\caption{The number of {\it future} major mergers that halos with initial ($z\approx2.2$) masses $10^{11.5}\Msun<M<10^{12.5}\Msun$ ({\it top}) and $10^{12.3}\Msun<M<10^{12.8}\Msun$ ({\it bottom}) undergo from $z\approx2.2$ to $z=0$. We distinguish between (a) ``main branch'' halos that undergo only minor mergers to $z=0$ ({\it filled blue column at $0$}), (b) ``main branch'' halos that undergo at least one major merger ({\it filled blue columns at $>0$}) and (c) ``disrupted'' halos ({\it red}). These are halos that at some stage are accreted onto a more massive halo in a minor merger event. For category (c), major mergers are counted only prior to merging with the larger halo. Around $60\%$ ($\approx70\%$) of the halos in the lower (higher) mass bin remain ``main branch'' halos. For both mass bins the mean number of future major mergers is $\approx1$, while fewer than $10\%$ undergo more than $2$ major mergers. These results do not depend on whether the initial halo at $z\approx2.2$ is identified as a major merger or on its dark matter accretion rate. We find that the mean number of major mergers a halo with mass $M$ undergoes between $z_i$ and $z_f<z_i$ is well approximated by: $\bar{N}_{\rm mm}(z_i,z_f,M)\approx0.13\times(\log(\frac{M}{10^{10}\tiny{\Msun}})+1)(z_i-z_f)$.}
\vspace{0.5cm}
\label{f:f3}
\end{figure}

A popular scenario is that the large central mass densities of SMGs and of $z\approx2$ compact red sequence galaxies are reduced by $z=0$ via dry dissipationless mergers \citep{vanDokkumP_05a,BellE_06b,NaabT_06a}. \citet{TacconiL_08a} show (in their Fig.~5) that this requires that the SMG masses grow by about an order of magnitude by $z=0$, assuming dry mergers with structurally similar systems, following \citet{NipotiC_03a}. We find that of the halos we have identified with the SMGs in \S\ref{s:z_2_smgs}, $\approx70\%$ remain ``main branch'' halos to $z=0$ (Fig.~\ref{f:f3}), and that their masses grow by factors of $\approx3-30$. This mass growth appears consistent with the requirement of the dry merger hypothesis. However, most of the mass growth does not occur via major mergers, since typically only $\approx1$ major merger occurs per halo to $z=0$, as shown by Fig.~\ref{f:f3}.

\begin{figure}[h]
\centering
\includegraphics[]{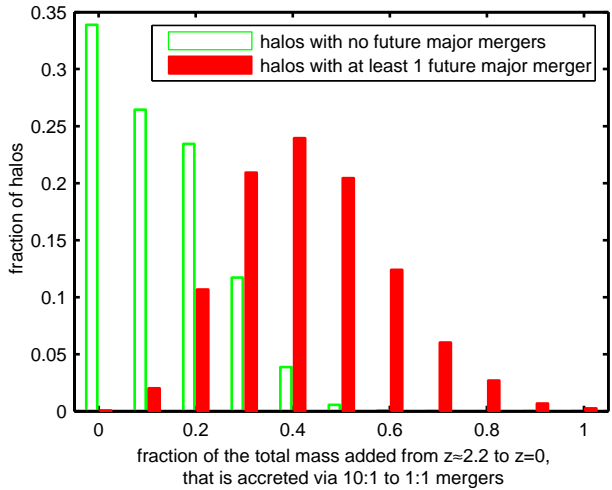}
\caption{Mass growth to $z=0$ of halos with initial ($z\approx2.2$) masses $10^{12.3}\Msun<M<10^{12.8}\Msun$ that remain "main branch" halos ({\it filled blue columns in the bottom Panel of Fig.~\ref{f:f3}}). The plot displays the relative number of halos versus the fraction of the total mass added from $z\approx2.2$ to $z=0$ that is accreted via mergers more intense than $10:1$. Results are shown for halos that undergo no major mergers to $z=0$ ({\it green}) and halos that undergo at least one major merger to $z=0$ ({\it filled red}). The results do not depend on whether the halo at $z\approx2.2$ is undergoing a major merger or on its accretion rate.}
\vspace{0.5cm}
\label{f:f4}
\end{figure}

Moreover, we show in Fig.~\ref{f:f4} that most of the mass growth is achieved via mergers less intense than $10:1$, which is qualitatively consistent with the idea that the growth of massive galaxies is not dominated by major mergers (e.g.~\citealp{HausmanM_78a,MallerA_06a,MasjediM_08a}). This is especially true for the halos that do not undergo further major mergers until $z=0$. Such halos tend to grow in mass only by a factor of $\approx3$, and gain $\gtrsim0.7$ of their new mass via mergers less intense than $10:1$. Also, the galaxies themselves probably grow even less than their dark matter halos. It seems unlikely that the mass accreted via such small halos can be sufficiently gas poor for minor dry mergers (e.g.~\citealp{BurkertA_07a}) to be an important growth mechanism (unless the galaxies are effectively stripped of their gas before merging with the descendant of the SMG \citep{NaabT_07a}).

A simpler and more likely explanation \citep{TacconiL_08a} is that the high SMG densities trace only the central starburst region, and exclude more extended fainter envelopes of pre-merger stars. At lower redshifts these stars would become visible and be of greater relative importance as the starburst fades, giving rise to larger half-light radii and smaller inferred densities (cf.~simulations by \citet{HopkinsP_08a}).

\section{Summary}
\label{s:summary}
We have constructed new halo merger trees from the \LCDM{ }Millennium Simulation. Our trees account for merger durations, and we use them to identify halos that are undergoing mergers and to extract dark matter accretion rates. We show that the high star formation rates observed in rotating disks at $z\approx2$ are plausibly consistent with the dark matter accretion rates expected for halos not undergoing major mergers. Given the measured star formation rates of SMGs, and the observationally supported assumption that they are undergoing major mergers, we infer their likely halo masses. Major mergers can neither lead to the complete transformation of the $z\approx2$ disks to $z=0$ ellipticals, nor to the disappearance of the very high density SMGs from $z\approx2$ to $z=0$. Therefore, secular/internal processes are likely important in the evolution of these high-redshift populations to present time.

\acknowledgements
We thank Gabriella De Lucia, Mike Boylan-Kolchin and Volker Springel for useful discussions, and the anonymous referee for valuable comments. The Millennium Simulation databases used in this paper and the web application providing online access to them were constructed as part of the activities of the German Astrophysical Virtual Observatory. We are grateful to Gerard Lemson who devotedly helped us to use the public databases. SG acknowledges the PhD fellowship of the International Max Planck Research School in Astrophysics, and the support received from a Marie Curie Host Fellowship for Early Stage Research Training.


\begin{thebibliography}{48}
\expandafter\ifx\csname natexlab\endcsname\relax\def\natexlab#1{#1}\fi


\bibitem[{{Bell} {et~al.}(2006){Bell}, {Naab}, {McIntosh}, {Somerville},
  {Caldwell}, {Barden}, {Wolf}, {Rix}, {Beckwith}, {Borch}, {H{\"a}ussler},
  {Heymans}, {Jahnke}, {Jogee}, {Koposov}, {Meisenheimer}, {Peng}, {Sanchez},
  \& {Wisotzki}}]{BellE_06b}
{Bell}, E.~F. {et~al.} 2006, \apj, 640, 241

\bibitem[{{Birnboim} \& {Dekel}(2003)}]{BirnboimY_03a}
{Birnboim}, Y., \& {Dekel}, A. 2003, \mnras, 345, 349

\bibitem[{{Blumenthal} {et~al.}(1984){Blumenthal}, {Faber}, {Primack}, \&
  {Rees}}]{BlumenthalG_84a}
{Blumenthal}, G.~R., {Faber}, S.~M., {Primack}, J.~R., \& {Rees}, M.~J. 1984,
  \nat, 311, 517

\bibitem[{{Bond} {et~al.}(1991){Bond}, {Cole}, {Efstathiou}, \&
  {Kaiser}}]{BondJ_91a}
{Bond}, J.~R., {Cole}, S., {Efstathiou}, G., \& {Kaiser}, N. 1991, \apj, 379,
  440

\bibitem[{{Bouch{\'e}} {et~al.}(2007){Bouch{\'e}}, {Cresci}, {Davies},
  {Eisenhauer}, {F{\"o}rster Schreiber}, {Genzel}, {Gillessen}, {Lehnert},
  {Lutz}, {Nesvadba}, {Shapiro}, {Sternberg}, {Tacconi}, {Verma}, {Cimatti},
  {Daddi}, {Renzini}, {Erb}, {Shapley}, \& {Steidel}}]{BoucheN_07a}
{Bouch{\'e}}, N. {et~al.} 2007, \apj, 671, 303

\bibitem[{{Bournaud} {et~al.}(2008){Bournaud}, {Daddi}, {Elmegreen},
  {Elmegreen}, {Nesvadba}, {Vanzella}, {di Matteo}, {Le Tiran}, {Lehnert}, \&
  {Elbaz}}]{BournaudF_08a}
{Bournaud}, F. {et~al.} 2008, \aap, 486, 741

\bibitem[{{Bower}(1991)}]{BowerR_91a}
{Bower}, R.~G. 1991, \mnras, 248, 332

\bibitem[{{Boylan-Kolchin} {et~al.}(2008){Boylan-Kolchin}, {Ma}, \&
  {Quataert}}]{Boylan-KolchinM_07a}
{Boylan-Kolchin}, M., {Ma}, C.-P., \& {Quataert}, E. 2008, \mnras, 383, 93

\bibitem[{{Burkert} {et~al.}(2007){Burkert}, {Naab}, \&
  {Johansson}}]{BurkertA_07a}
{Burkert}, A., {Naab}, T., \& {Johansson}, P.~H. 2007, \apj, submitted
  (astro-ph/07100663)

\bibitem[{{Conroy} {et~al.}(2008){Conroy}, {Shapley}, {Tinker}, {Santos}, \&
  {Lemson}}]{ConroyC_08a}
{Conroy}, C., {Shapley}, A.~E., {Tinker}, J.~L., {Santos}, M.~R., \& {Lemson},
  G. 2008, \apj, 679, 1192

\bibitem[{{Conselice}(2003)}]{ConseliceC_03b}
{Conselice}, C.~J. 2003, \apjs, 147, 1

\bibitem[{{Conselice}(2006)}]{ConseliceC_06a}
---. 2006, \apj, 638, 686

\bibitem[{{Daddi} {et~al.}(2007){Daddi}, {Dickinson}, {Morrison}, {Chary},
  {Cimatti}, {Elbaz}, {Frayer}, {Renzini}, {Pope}, {Alexander}, {Bauer},
  {Giavalisco}, {Huynh}, {Kurk}, \& {Mignoli}}]{DaddiE_07a}
{Daddi}, E. {et~al.} 2007, \apj, 670, 156

\bibitem[{{Davis} {et~al.}(1985){Davis}, {Efstathiou}, {Frenk}, \&
  {White}}]{DavisM_85a}
{Davis}, M., {Efstathiou}, G., {Frenk}, C.~S., \& {White}, S.~D.~M. 1985, \apj,
  292, 371

\bibitem[{{Fakhouri} \& {Ma}(2008)}]{FakhouriO_07a}
{Fakhouri}, O., \& {Ma}, C.-P. 2008, \mnras, 359

\bibitem[{{F{\"o}rster Schreiber} {et~al.}(2006){F{\"o}rster Schreiber},
  {Genzel}, {Lehnert}, {Bouch{\'e}}, {Verma}, {Erb}, {Shapley}, {Steidel},
  {Davies}, {Lutz}, {Nesvadba}, {Tacconi}, {Eisenhauer}, {Abuter}, {Gilbert},
  {Gillessen}, \& {Sternberg}}]{FoersterSchreiberN_06a}
{F{\"o}rster Schreiber}, N.~M. {et~al.} 2006, \apj, 645, 1062

\bibitem[{{Genzel} {et~al.}(2008){Genzel}, {Burkert}, {Bouche}, {Cresci},
  {Foerster Schreiber}, {Shapley}, {Shapiro}, {Tacconi}, {Buschkamp},
  {Cimatti}, {Daddi}, {Davies}, {Eisenhauer}, {Erb}, {Genel}, {Gerhard},
  {Hicks}, {Lutz}, {Naab}, {Ott}, {Rabien}, {Renzini}, {Steidel}, {Sternberg},
  \& {Lilly}}]{GenzelR_08a}
{Genzel}, R. {et~al.} 2008, \apj, accepted (astro-ph/08071184)

\bibitem[{{Genzel} {et~al.}(2006){Genzel}, {Tacconi}, {Eisenhauer},
  {F{\"o}rster Schreiber}, {Cimatti}, {Daddi}, {Bouch{\'e}}, {Davies},
  {Lehnert}, {Lutz}, {Nesvadba}, {Verma}, {Abuter}, {Shapiro}, {Sternberg},
  {Renzini}, {Kong}, {Arimoto}, \& {Mignoli}}]{GenzelR_06a}
---. 2006, \nat, 442, 786

\bibitem[{{Hausman} \& {Ostriker}(1978)}]{HausmanM_78a}
{Hausman}, M.~A., \& {Ostriker}, J.~P. 1978, \apj, 224, 320

\bibitem[{{Hernquist} \& {Mihos}(1995)}]{HernquistL_95a}
{Hernquist}, L., \& {Mihos}, J.~C. 1995, \apj, 448, 41

\bibitem[{{Hopkins} {et~al.}(2008{\natexlab{a}}){Hopkins}, {Cox}, {Kere{\v s}},
  \& {Hernquist}}]{HopkinsP_07b}
{Hopkins}, P.~F., {Cox}, T.~J., {Kere{\v s}}, D., \& {Hernquist}, L.
  2008{\natexlab{a}}, \apjs, 175, 390

\bibitem[{{Hopkins} {et~al.}(2008{\natexlab{b}}){Hopkins}, {Hernquist}, {Cox},
  {Dutta}, \& {Rothberg}}]{HopkinsP_08a}
{Hopkins}, P.~F., {Hernquist}, L., {Cox}, T.~J., {Dutta}, S.~N., \& {Rothberg},
  B. 2008{\natexlab{b}}, \apj, 679, 156

\bibitem[{{Jiang} {et~al.}(2008){Jiang}, {Jing}, {Faltenbacher}, {Lin}, \&
  {Li}}]{JiangC_07a}
{Jiang}, C.~Y., {Jing}, Y.~P., {Faltenbacher}, A., {Lin}, W.~P., \& {Li}, C.
  2008, \apj, 675, 1095

\bibitem[{{Kere\v{s}} {et~al.}(2005){Kere\v{s}}, {Katz}, {Weinberg}, \&
  {Dav{\'e}}}]{KeresD_05a}
{Kere\v{s}}, D., {Katz}, N., {Weinberg}, D.~H., \& {Dav{\'e}}, R. 2005, \mnras,
  363, 2

\bibitem[{{Law} {et~al.}(2007){Law}, {Steidel}, {Erb}, {Larkin}, {Pettini},
  {Shapley}, \& {Wright}}]{LawD_07a}
{Law}, D.~R., {Steidel}, C.~C., {Erb}, D.~K., {Larkin}, J.~E., {Pettini}, M.,
  {Shapley}, A.~E., \& {Wright}, S.~A. 2007, \apj, 669, 929

\bibitem[{{Lotz} {et~al.}(2008){Lotz}, {Jonsson}, {Cox}, \&
  {Primack}}]{LotzJ_08b}
{Lotz}, J.~M., {Jonsson}, P., {Cox}, T.~J., \& {Primack}, J.~R. 2008, \mnras,
  submitted (astro-ph/08051246)

\bibitem[{{Maller} {et~al.}(2006){Maller}, {Katz}, {Kere{\v s}}, {Dav{\'e}}, \&
  {Weinberg}}]{MallerA_06a}
{Maller}, A.~H., {Katz}, N., {Kere{\v s}}, D., {Dav{\'e}}, R., \& {Weinberg},
  D.~H. 2006, \apj, 647, 763

\bibitem[{{Masjedi} {et~al.}(2008){Masjedi}, {Hogg}, \&
  {Blanton}}]{MasjediM_08a}
{Masjedi}, M., {Hogg}, D.~W., \& {Blanton}, M.~R. 2008, \apj, 679, 260

\bibitem[{{Naab} \& {Burkert}(2003)}]{NaabT_03a}
{Naab}, T., \& {Burkert}, A. 2003, \apj, 597, 893

\bibitem[{{Naab} {et~al.}(2007){Naab}, {Johansson}, {Ostriker}, \&
  {Efstathiou}}]{NaabT_07a}
{Naab}, T., {Johansson}, P.~H., {Ostriker}, J.~P., \& {Efstathiou}, G. 2007,
  \apj, 658, 710

\bibitem[{{Naab} {et~al.}(2006){Naab}, {Khochfar}, \& {Burkert}}]{NaabT_06a}
{Naab}, T., {Khochfar}, S., \& {Burkert}, A. 2006, \apjl, 636, L81

\bibitem[{{Neistein} {et~al.}(2006){Neistein}, {van den Bosch}, \&
  {Dekel}}]{NeisteinE_06a}
{Neistein}, E., {van den Bosch}, F.~C., \& {Dekel}, A. 2006, \mnras, 372, 933

\bibitem[{{Nipoti} {et~al.}(2003){Nipoti}, {Londrillo}, \&
  {Ciotti}}]{NipotiC_03a}
{Nipoti}, C., {Londrillo}, P., \& {Ciotti}, L. 2003, \mnras, 342, 501

\bibitem[{{Noeske} {et~al.}(2007){Noeske}, {Weiner}, {Faber}, {Papovich},
  {Koo}, {Somerville}, {Bundy}, {Conselice}, {Newman}, {Schiminovich}, {Le
  Floc'h}, {Coil}, {Rieke}, {Lotz}, {Primack}, {Barmby}, {Cooper}, {Davis},
  {Ellis}, {Fazio}, {Guhathakurta}, {Huang}, {Kassin}, {Martin}, {Phillips},
  {Rich}, {Small}, {Willmer}, \& {Wilson}}]{NoeskeK_07a}
{Noeske}, K.~G. {et~al.} 2007, \apjl, 660, L43

\bibitem[{{Ocvirk} {et~al.}(2008){Ocvirk}, {Pichon}, \&
  {Teyssier}}]{OcvirkP_08a}
{Ocvirk}, P., {Pichon}, C., \& {Teyssier}, R. 2008, \mnras, submitted
  (astro-ph/08034506)

\bibitem[{{Press} \& {Schechter}(1974)}]{PressW_74a}
{Press}, W.~H., \& {Schechter}, P. 1974, \apj, 187, 425

\bibitem[{{Shapiro} {et~al.}(2008){Shapiro}, {Genzel}, {Forster Schreiber},
  {Tacconi}, {Bouche}, {Cresci}, {Davies}, {Eisenhauer}, {Johansson},
  {Krajnovic}, {Lutz}, {Naab}, {Arimoto}, {Arribas}, {Cimatti}, {Colina},
  {Daddi}, {Daigle}, {Erb}, {Hernandez}, {Kong}, {Mignoli}, {Onodera},
  {Renzini}, {Shapley}, \& {Steidel}}]{ShapiroK_08a}
{Shapiro}, K.~L. {et~al.} 2008, \apj, accepted (astro-ph/08020879)

\bibitem[{{Springel} {et~al.}(2006){Springel}, {Frenk}, \&
  {White}}]{SpringelV_06a}
{Springel}, V., {Frenk}, C.~S., \& {White}, S.~D.~M. 2006, \nat, 440, 1137

\bibitem[{{Springel} \& {Hernquist}(2005)}]{SpringelV_05b}
{Springel}, V., \& {Hernquist}, L. 2005, \apjl, 622, L9

\bibitem[{{Springel} {et~al.}(2005){Springel}, {White}, {Jenkins}, {Frenk},
  {Yoshida}, {Gao}, {Navarro}, {Thacker}, {Croton}, {Helly}, {Peacock}, {Cole},
  {Thomas}, {Couchman}, {Evrard}, {Colberg}, \& {Pearce}}]{SpringelV_05a}
{Springel}, V. {et~al.} 2005, \nat, 435, 629

\bibitem[{{Springel} {et~al.}(2001){Springel}, {White}, {Tormen}, \&
  {Kauffmann}}]{SpringelV_01}
{Springel}, V., {White}, S.~D.~M., {Tormen}, G., \& {Kauffmann}, G. 2001,
  \mnras, 328, 726

\bibitem[{{Steinmetz} \& {Navarro}(2002)}]{SteinmetzM_02a}
{Steinmetz}, M., \& {Navarro}, J.~F. 2002, New Astronomy, 7, 155

\bibitem[{{Tacconi} {et~al.}(2008){Tacconi}, {Genzel}, {Smail}, {Neri},
  {Chapman}, {Ivison}, {Blain}, {Cox}, {Omont}, {Bertoldi}, {Greve},
  {F{\"o}rster Schreiber}, {Genel}, {Lutz}, {Swinbank}, {Shapley}, {Erb},
  {Cimatti}, {Daddi}, \& {Baker}}]{TacconiL_08a}
{Tacconi}, L.~J. {et~al.} 2008, \apj, 680, 246

\bibitem[{{Tacconi} {et~al.}(2006){Tacconi}, {Neri}, {Chapman}, {Genzel},
  {Smail}, {Ivison}, {Bertoldi}, {Blain}, {Cox}, {Greve}, \&
  {Omont}}]{TacconiL_06a}
---. 2006, \apj, 640, 228

\bibitem[{{Toomre}(1977)}]{ToomreA_77a}
{Toomre}, A. 1977, in Evolution of Galaxies and Stellar Populations, 401

\bibitem[{{van Dokkum}(2005)}]{vanDokkumP_05a}
{van Dokkum}, P.~G. 2005, \aj, 130, 2647

\bibitem[{{van Dokkum} {et~al.}(2008){van Dokkum}, {Franx}, {Kriek}, {Holden},
  {Illingworth}, {Magee}, {Bouwens}, {Marchesini}, {Quadri}, {Rudnick},
  {Taylor}, \& {Toft}}]{vanDokkumP_07a}
{van Dokkum}, P.~G. {et~al.} 2008, \apjl, 677, L5

\bibitem[{{van Starkenburg} {et~al.}(2008){van Starkenburg}, {van der Werf},
  {Franx}, {Labbe}, {Rudnick}, \& {Wuyts}}]{vanStarkenburg_08a}
{van Starkenburg}, L., {van der Werf}, P.~P., {Franx}, M., {Labbe}, I.,
  {Rudnick}, G., \& {Wuyts}, S. 2008, \aap, accepted (astro-ph/08063369)


\end{thebibliography}
\end{document}